\documentstyle[prl,aps,epsfig,preprint]{revtex}

\def \lacuo{La$_2$CuO$_4$}
\def \ndcuo{Nd$_2$CuO$_4$}
\def \prcuo{Pr$_2$CuO$_4$}

\def \ybacuo{YBa$_2$Cu$_3$O$_{6+x}$}

\begin{document}
\draft

\title {{\bf Superexchange coupling and  spin susceptibility spectral 
weight in undoped monolayer cuprates }}

\author{P. Bourges$^{(1)}$, H. Casalta$^{(1,2)}$, 
A.S. Ivanov$^{(2,3)}$ and D. Petitgrand$^{(1)}$ 
}

\address{1 - Laboratoire L\'eon Brillouin, CEA-CNRS, CE Saclay, 91191 Gif sur Yvette, France}
\address{2 - Institut Laue-Langevin, 156X, 
38042 Grenoble Cedex 9, France}
\address{3 -  Kurchatov Institute of Atomic Energy, 123182 Moscow, Russia}

\maketitle

\begin{abstract}

 A systematic inelastic neutron scattering  study  of  the  superexchange 
interaction in three different undoped monolayer  cuprates (\lacuo, \ndcuo\ and 
\prcuo) has been performed  using conventional  triple axis technique. 
We deduce the in-plane antiferromagnetic (AF) superexchange 
coupling $J$ which actually presents no simple relation versus 
crystallographic parameters. 
The absolute spectral weight of the spin susceptibility has been 
obtained and it is found to be  smaller than expected even when quantum 
corrections of the AF ground state are taken into account. 
\end{abstract}

 % ]

The copper spins properties  of the insulating cuprates are of 
particular interest as they give insights into the microscopic 
description of the high-$T_C$ superconductors. Undoped parent compounds 
of many high-$T_C$ 
cuprates are usually described as Mott-Hubbard insulators. They exhibit
an antiferromagnetic ordering  below a N\'eel temperature 
 ranging between 250 K and 420 K. This N\'eel state is well 
accounted for by a spin-${1\over2}$ antiferromagnetic (AF) Heisenberg model 
on a square lattice\cite{manou}. The following Hamiltonian, 
$H=-J \sum_{<ij>}S_i S_j$ where the sum is performed over spin pairs, is
then used to describe the AF ground state where the  most essential and 
generic parameter is the huge Cu-O-Cu superexchange 
interaction, $J$, within the $CuO_2$ plane. $J$ is usually determined by
inelastic neutron scattering (INS) experiments which probe the dispersion
relations of spin-wave excitations. 
The intraplane AF superexchange is then deduced from the measured 
spin-wave velocity $c$, as $c= 2 S {\sqrt 2} Z_c J a$ (where $a$ is the square 
lattice constant, S=${1\over2}$ and $Z_c\simeq 1.18$ represents 
quantum corrections of the 
AF ground state). Unfortunately, due to the large steepness of the 
in-plane spin wave dispersion (related to the  large value of  J $\approx$ 
100-150 meV), the spin-wave velocity is not easily deduced from INS experiments.
Therefore, a precise knowledge of  $J$ is still needed in parent compounds 
of cuprates. Another essential magnetic parameter is the spectral weight
of copper spin susceptibility which has been, so far, only reported  
in \lacuo\cite{itoh,hayden}. The importance of these two parameters 
has been recently emphasized in doped
materials as $J$ is found to be renormalized compared to the undoped 
case and the spectral weight is shifted to lower energy\cite{hayden,bourges}.

Here, we present, by systematic neutron scattering measurements, the
spin wave excitations of three different parent compounds of single-$CuO_2$ 
layer cuprates. Especially, using an adapted  focalised neutron scattering 
geometry, we are able to determine their spin-velocity with accuracy and 
to deduce $J$. Furthermore, we have determined the spectral weight of the
spin susceptibility in absolute units and the perpendicular
spin susceptibility, $\chi_{\perp}$. $\chi_{\perp}$ can be also obtained 
as a consequence of sum-rules by applying the hydrodynamics relation, 
$\rho_s=(c/a)^2\chi_{\perp}$\cite{manou,chn}, where $\rho_s$ is the 
spin-stiffness constant. We found that $\chi_{\perp}$ measured in 
neutron experiments is smaller than expected from this relation.
This reduction of about 30\%  is presumably due to covalent effects
between copper d-orbitals and  oxygen p-orbitals.
% The quantum corrections associated with  $\rho_s$, and $\chi_{\perp}$ 
% have been experimentally deduced. In contrast to usual knowledge, they 
%are found   to differ from one single layer cuprate to the other. 
% We report $J$ in different cuprates in relation  with their structural 
% properties. 

High quality   \lacuo, \ndcuo\, and \prcuo\ single crystals of similar volume  
of about 0.5 cm$^3$ were used. Neodymium and Prasedymium-based samples exhibit 
a N\'eel temperature around 250  K whereas the AF transition 
occurs just above room temperature, 320 K, in the Lanthanum-based 
sample\cite{helene}. The samples were mounted with the reciprocal 
directions (110) and (001)  within the scattering plane 
[these directions are referring to the tetragonal reciprocal lattice
 with $Q=(h,h,q_c)$. We used the same axis in the case of 
orthorhombic \lacuo ]. Inelastic neutron scattering has been performed on the
triple axis spectrometers 1T and  4F1, installed  respectively on thermal 
and cold source beams at the Orph\'ee reactor, Saclay. The (002) reflection
of Pyrolytic Graphite was used for both monochromator and analyser. No collimation
was used and a filter (Graphite one on 1T and Beryllium one on 4F1)
was placed on the scattered beam to remove higher order contaminations.

A special scattering geometry\cite{shamoto} was used in order to align 
the resolution spectrometer ellipsoid along the AF line, i.e. the (001) 
direction.  Namely, this focalisation allows us to separate 
counterpropagating spin-waves at relatively low energies as compared 
with standard geometries\cite{rossat,aeppli}. We extend this technique 
down to 15 meV.  For such a geometry, only one 
$q_c$ value is accessible  for a fixed energy transfer and a fixed 
final neutron energy. To be powerful, this geometry also requires very 
good sample mosaicities.

\begin{figure}

\centerline{\epsfig{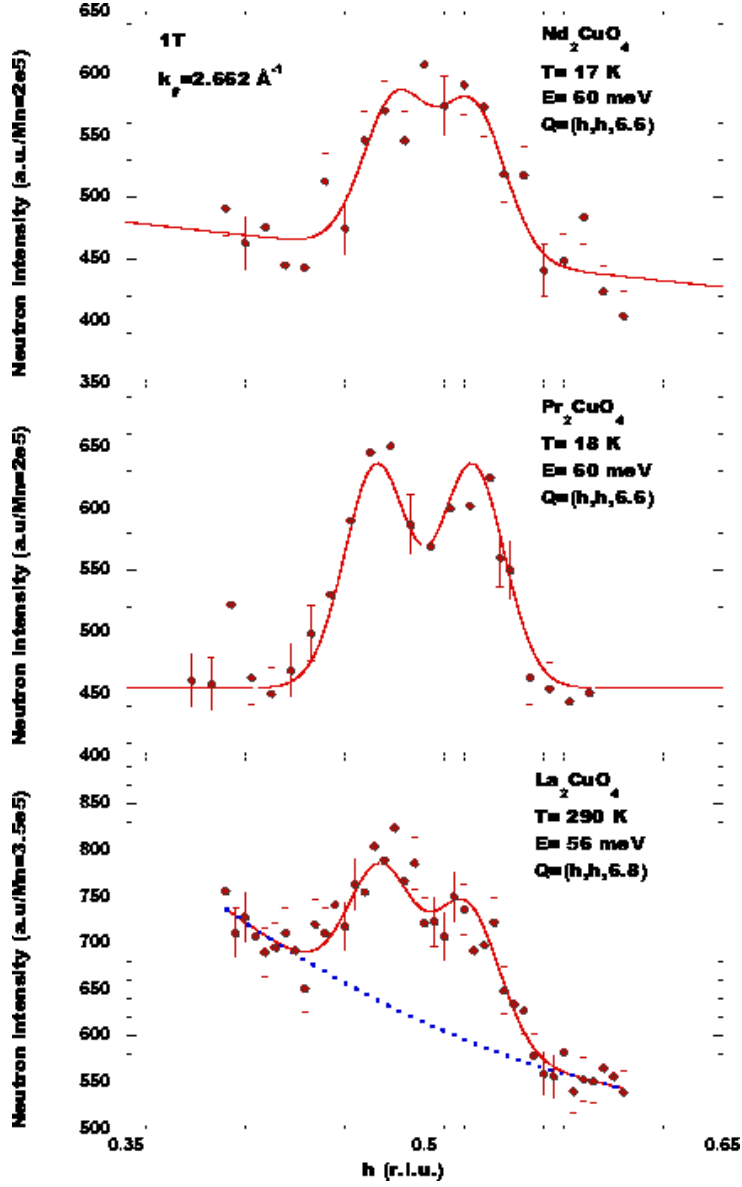}}
\caption{q-scans across the magnetic line around $\hbar\omega \simeq$
60 meV in three different monolayer undoped cuprates. Typical counting
time is 1 hour per point. 
Full lines correspond to the convolution product of the 
Gaussian resolution ellipsoid by the spin-wave cross section 
(\ref{sqw}) with the spin susceptibility (\ref{chisec}). 
% The scans have been renormalized to the sample volume.
 }
\label{fig60mev} \end{figure}

\begin{figure}
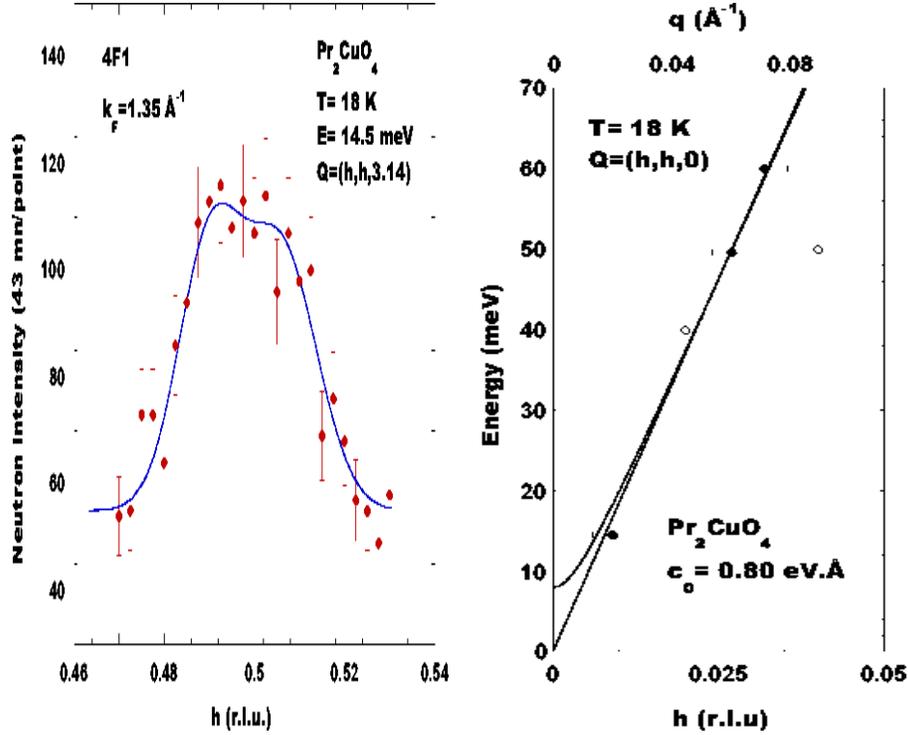

\begin{tabular}{cc}
\epsfig{file=fig14mev.epsi,height=10 cm,width=6 cm} & 
\epsfig{file=figdisp.epsi,height=10 cm,width=6 cm} \\
\end{tabular}
\caption[figdisp]{Left: q-scan across the magnetic line at $\hbar\omega=$
14.5 meV in \prcuo\ (see Fig. \ref{fig60mev} for details). Right: 
In-plane magnon dispersion in \prcuo.  At low energy, 
the degeneracy between out-of-plane and in-plane spin components
is removed due to planar anisotropy leading to an  out-of-plane spin  gap
of about 8 meV\cite{ivanov}.  Above $\sim$ 12 meV, both spin 
components become very rapidly indistinguishable with increasing the energy.
Open circles correspond to a previous measurement\cite{sumarlin}. }
\label{figdisp} \end{figure}

We now present q-scans (constant energy transfer scan) along the (110)
direction in the three different monolayer cuprates: \lacuo, \ndcuo\ and 
\prcuo. Figure \ref{fig60mev} depicts q-scans measured at  an
energy transfer around 60 meV using the 
same experimental setup. The double peak structure is clearly seen in 
\lacuo\ and in \prcuo\ whereas only a flattened peak shape is observed in 
\ndcuo. This difference emphasizes a larger spin velocity in \ndcuo. 
In order to improve at low energy the determination of the spin velocity, 
we have applied in \prcuo\  this focalised  geometry down to 
$E=14.5$ meV, where a flattened peak shape is found 
(Fig. \ref{figdisp}). Our data in \prcuo\ represent a clear improvement 
of a previous measurement\cite{sumarlin}.

Here, we focus on the low energy part of the spin wave spectrum 
in the limit where the dispersion relation for AF magnons is 
linear ($\hbar\omega << 2 Z_c J$). However, at low energy, the 
magnon spectrum exhibits gaps which are either related to 
planar anisotropy or to  interlayer interactions\cite{rossat}. 
The usual linear relation is thus only recovered  for
energies  slightly larger than these gaps. Due to the large 
intraplane superexchange interaction  in cuprates, this condition is 
fulfilled for energy above $\sim$ 12 meV (see Fig.  \ref{figdisp}).
Above this energy, the spin-wave neutron cross section per formula unit 
can be written in terms of the dynamical spin 
susceptibility\cite{lovesey,note}, $\chi({\bf Q},\omega)$, as 
\begin{equation}
{{d^2\sigma}\over{d\Omega d\omega}} = r_0^2 % \frac{k_F}{k_I}\;
 \frac{F^2({\bf Q})}{\pi (g \mu_B)^2}  {1\over2}(3-\frac{Q_c^2}{Q^2})
 \frac{Im \chi({\bf Q},\omega)}{1 - \exp(-\hbar\omega/k_B T)}
\label{sqw}
\end{equation}
where  $r_0^2$=0.292 barns, 
% $k_I$ ($k_F$) is the incident (final) neutron wavevecto,
 $F({\bf Q})$ is the atomic form factor of  Cu$^{2+}$ 
spins\cite{formfactor}, $g\approx 2$ is the Land\'e factor for copper spins,
and $Q_c=\frac{2 \pi}{c} q_c$ is the component 
along the (001) direction of the scattered wavevector, ${\bf Q}$.
For an AF single layer cuprate, the imaginary part of dynamical 
susceptibility of the low energy spin wave excitations 
is  given in absolute units by\cite{note}

\begin{equation}
Im\chi ({\bf Q}, \omega) \; =  S \pi Z_\chi Z_c (g \mu_B)^2\; 
\frac{\sqrt{2}}{q a} \delta[\omega - c q]
\label{chisec}
\end{equation}
where $q$ is the in-plane wavevector component  along the (110) direction  
referred to the AF wavevector. The quantum corrections associated to 
the perpendicular susceptibility\cite{manou}, $Z_{\chi}$, is included.
The convolution product of the Gaussian resolution ellipsoid by 
the spin-wave cross section (\ref{sqw}) with the spin susceptibility 
(\ref{chisec}) gives i) the dispersion relation of  magnons ii)
the spectral weight of $Im\chi$. The q-scans have been fitted by this
convolution product with 4 fitting parameters:
the magnon in-plane wavector $q$, the amplitude of $Im\chi$ and a sloping background. We note that the observed experimental q-width along the 
(110) direction merely corresponds to that of the resolution. 

In \prcuo, the in-plane magnon dispersion is reported in Fig. 
\ref{figdisp} over a wide energy range. As  expected, 
a linear  dispersion typical of AF excitations is found with a 
slope which is the spin wave velocity, $c$= 0.80 eV.\AA.
Comparison of the different q-scans (fig. \ref{fig60mev}) gives  0.85 eV.\AA\ 
for \lacuo\ in agreement with a previous determination by high energy 
neutron experiments\cite{aeppli} and $c$=1.02  eV.\AA\ for \ndcuo\
(see Table (\ref{table1})). The  magnon wavevector, and so the spin velocity 
and the AF intraplane superexchange, are then found  larger 
for \ndcuo\ by about  20\% as compared with the two other systems. 

The spin susceptibility in absolute units has been experimentally 
estimated by a standard calibration\cite{bourges} using acoustic phonons, 
whose dynamical structure factor is known by lattice dynamics.
 The magnetic part has been measured 
from high energy scans (Fig. \ref{fig60mev}) as well as non-resolved low 
energy q-scans. In order to compare the observed spin susceptibility 
in absolute units with its theoretical predictions\cite{manou}, we 
calculate the average of (\ref{chisec}) over the two dimensional (2D) 
q-space perpendicular to the (001) direction, 
$\tilde{\chi}_{2D}= \int d {\bf q}_{\rm 2D} Im\chi ({\bf Q},\omega) 
/ \int d {\bf q}_{\rm 2D}$. In our experimental energy range,
$\tilde{\chi}_{2D}$ is almost independent of energy:
$\tilde{\chi}_{2D}\simeq S (g \mu_B)^2 Z_\chi/2 J$. % \cite{calibre}. 
Values for $\tilde{\chi}_{2D}$ are listed in Table (\ref{table1}).
In \lacuo, it compares well with two previous measurements\cite{itoh,hayden}. 
On the one hand, Itoh {\it et al.}\cite{itoh} have reported an effective value 
of S=0.17 which is reduced from the spin number, S=1/2. That agrees with our 
observed spin susceptibility, 2.7  $\mu_B^2$/eV 
(see Table (\ref{table1})), which is reduced by the same factor from 
the classical spin susceptibility (without quantum corrections), 
$\tilde{\chi}^{class}_{2D}\simeq S (g \mu_B)^2 / 2 J = 7.5\ \mu_B^2$/eV.
On the other hand,  Hayden {\it et al.}\cite{hayden} have obtained $\tilde{\chi}_{2D}= 2.5\ \mu_B^2$/eV which agrees in errors with our 
value\cite{notehayden}.

The perpendicular susceptibility, $\chi_{\perp}$,
deduced from our INS measurements is then obtained by applying the relation 
$\chi_{\perp}= \tilde{\chi}_{2D}/4 S (g \mu_B)^2$\cite{manou} % ,calibre} 
and listed in Table (\ref{table1}). $\chi_{\perp}$ can be independently 
deduced from the spin stiffness, $\rho_s$, applying standard 
hydrodynamics relation in the Heisenberg model
(see Table (\ref{table1})). Let us recall that the spin-stiffness constant 
has been obtained in the Heisenberg model from the two-dimensional
correlation length $\xi_{2D}$ above the N\'eel temperature as, 
$\xi_{2D} \propto \exp(\frac{2 \pi \rho_s}{ k_B T})$\cite{chn},
$\xi_{2D}$ being itself measured using energy integrated neutron 
scattering\cite{2axes,thurston}. Surprisingly, the value of $\chi_{\perp}$
deduced from $\rho_s$ is found to be systematically smaller than the
one measured in INS experiments even when quantum corrections of the 
AF ground state are considered. This discrepancy of about 32\% for 
the three compounds is likely due to the covalence of copper d-orbitals 
with oxygen p-orbitals\cite{covalence}.  Reducing the absolute scale
of the atomic form factor, such effects can explain the 
diminution of the inelastic spectral weight of the spin susceptibility
as well as the low temperature ordered magnetization value\cite{rossat}.
Consequently, the spectral weight of $Im\chi$ does not solely
determine the quantum corrections for the spin susceptibility.

We now deduce $J$ as well as the quantum corrections. Since there are
more unknown parameters that the measured ones, we need to use 
theoretical estimation for one parameter. Among the measured magnetic 
parameters, the spin wave dispersion curve is presumably the less 
altered by frustration effect and disorder\cite{singhhuse}. 
The quantum correction to the spin wave velocity $Z_c$
estimated  from different theoretical approaches\cite{manou,singhhuse} 
likely converges to a best value of $Z_c= 1.18$\cite{igarashi}.  $J$ is then 
confidently deduced from the spin wave velocity using this
value (see Table (\ref{table1})). Two other parameters are related to 
$J$. On the one hand, the spin-stiffness constant is usually 
modelled as $ \rho_s = Z_{\rho_s} J S^2$\cite{singhhuse}
(where $Z_{\rho_s}$ accounts for quantum effects).
On the other hand,  a high frequency broad peak is observed in 
Raman scattering which is likely interpreted as two-magnons processes with 
opposite momenta\cite{sulewski,tomeno}. By means of series expansions 
technique\cite{singh}, the moments of the Raman intensity (the frequency 
of the spectrum maxima $M_1$ as well as lineshapes) have been related 
to $J$, for instance $M_1/J=3.58$. The quantum corrections for the 
spin stiffness $Z_{\rho_s}$, the perpendicular susceptibility $Z_\chi$,
and the ratio between the first Raman scattering moment and  $J$
have been obtained and also listed in Table (\ref{table1}).

% In contrast to what is generally accepted, these parameters in Table 
% (\ref{table1}) differ significantly from one system to the other.
Surprisingly, only the quantum corrections found in \lacuo\ are in agreement 
with the theoretical predictions\cite{manou} either based 
on series expansions\cite{singhhuse,singh} or  based on 
1/S expansion linear spin-wave theory\cite{igarashi}: 
$Z_{\rho_s}= 0.72$ and $Z_\chi = 0.51$ and $\omega_R/J$=3.58. 
The two other systems display larger quantum 
corrections for $\rho_s$ and $\chi_{\perp}$ may be related to their 
different low energy spin excitations\cite{ivanov}. An even larger 
discrepancy is observed for the spin pair Raman scattering 
measurements. Consequently, the neutron measurements which 
determine $\rho_s$ as well as the light scattering experiments 
only give a rough estimation of $J$.
% , via the temperature dependence of the correlation length,

\begin{figure}
\centerline{\epsfig{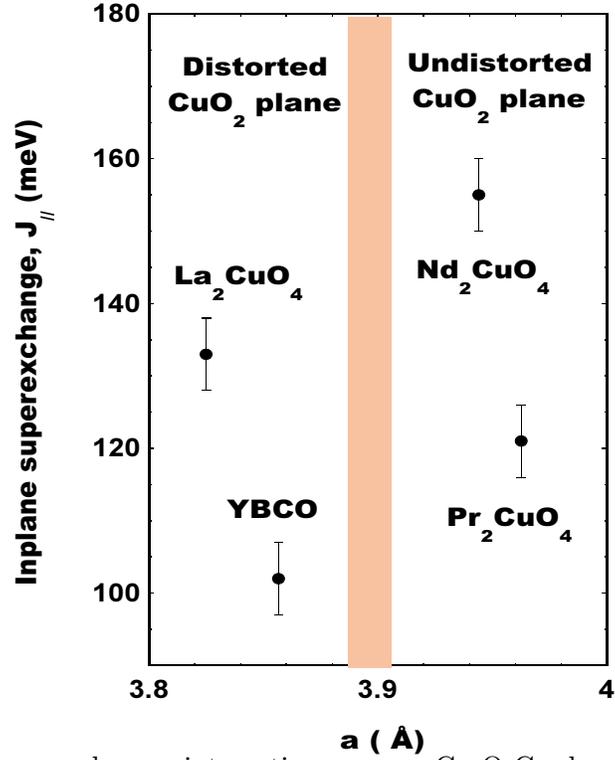}}
\caption[figJfcucu]{In-plane superexchange interaction versus 
 Cu-O-Cu bonding length in different cuprates. The value for 
the bilayer system YBCO is from \cite{shamoto}.}
\label{figJfcucu} \end{figure}

We now relate the copper spin intraplane  superexchange 
determined by INS with the crystallographic distances between copper 
atoms (Figure \ref{figJfcucu}). Clearly,  $J$ does not exhibit 
a monotonous dependence versus the bonding Cu-O-Cu length in contrast
to what could be expected. This outlines that the classical superexchange 
theory being only related to the Cu-O-Cu bonding is a too simple description. 
Moreover, it has been recently stressed that the large enhancement of $J$ 
is actually caused by another structural unit, namely the Cu-O-O 
triangle\cite{eskes}.  Empirically, one can distinguish 
distorted tetragonal lattice and perfect square one. Indeed, 
$J$ appears to decrease sharply 
with the distances between copper atoms in \ndcuo\ and in \prcuo\
(both having the T'-phase, i.e. linear Cu-O-Cu bonding). 
Note that the largest  
$J$ is found in \ndcuo\ where the Cu-O distance exactly 
corresponds to the sum of copper and oxygen ionic radius.
The two other systems do not belong to the same family as the bonding 
Cu-O-Cu  is not linear: it is distorted perpendicular to the plane in \ybacuo\cite{rossat}, or even in both directions in  \lacuo\cite{braden} 
due to the tilt of the $CuO_6$ octahedra. Therefore, $J$ turns out to be 
extremely sensitive function of Cu-O-Cu bonding angle. 

In conclusion, by means of  inelastic neutron scattering 
experiments using conventional triple-axis technique, we deduce
$J$ and the quantum corrections of the AF ground state  
in undoped  monolayer cuprates.
The in-plane antiferromagnetic superexchange coupling $J$ does 
not exhibit a monotonous behaviour versus the bonding Cu-O-Cu length. 
The absolute spectral weight of the spin susceptibility is 
smaller than expected from quantum corrections\cite{manou}, likely 
due to covalent effects. These results provide a necessary ground  
for the understanding of antiferromagnetism 
in the high-$T_C$ superconductors.

\noindent {\bf Acknowledgments}\\

We wish to thank  S. Aubry.  G. Collin, B. Hennion,   S.V. Maleyev, and
L.P. Regnault for stimulating discussions. We also acknowledge 
L. Pintschovius and M. Braden for their help concerning the \lacuo\ sample.

\clearpage

\begin{table}[h]
\begin{tabular}{c c c c c c c c c c c}
% \hline
Parameter & $T_N$&$c$ & $\tilde{\chi}_{2D}$ $\Rightarrow$ & 
$\chi_{\perp}$(INS)& $2\pi\rho_s$ $\Rightarrow$& $\chi_{\perp}$ 
& $J$ & $Z_{\rho_s}$ & $Z_{\chi}=Z_{\rho_s}/Z^2_c$ & 
$\omega_R/J$\\
Units &K&meV\AA& $\mu_B^2$/eV& eV$^{-1}$ & meV & eV$^{-1}$& meV& &\\
\hline
Errors & & $\pm$ 20 &  $\pm$ 0.4& $\pm$ 0.05& $\pm$ 5 & $\pm$ 0.04 
& $\pm$ 3& $\pm$ 0.05 & $\pm$ 0.04\\
\hline
\lacuo & 320 &850 & 2.7 &  0.34& 150$^a$ & 0.48 & 133 & 0.72 & 0.52 &  3.5$^c$\\
\ndcuo & 246 & 1020& 1.8 & 0.22 & 137$^b$ & 0.33 & 155 & 0.64 & 0.46 &  2.5$^c$\\
\prcuo & 252 & 800 & 2.3& 0.29&114$^b$ & 0.44 & 121 & 0.6  & 0.43 &  3.1$^d$\\
% \hline
%theory &  & $Z_c J \sqrt{2} a$ & $(g\mu_B)^2 Z_{\chi}/4 J$ & $\pi Z_{\rho_s} J/2$ & $\rho_s (a/c)^2$&  & 0.71 & 0.52 &  3.58\\
% \hline \hline
\end{tabular}
\caption[table1] {Magnetic parameters in three undoped single layer  cuprates.
The value of the spin stiffness has been deduced from previous 
energy-integrated neutron scattering experiments: $^a$ from \cite{2axes}, $^b$ from \cite{thurston,sumarlin}.  $\omega_R$ is the first moment of 
the Raman scattering data: $^c$ from \cite{sulewski}, $^d$ from \cite{tomeno}.
Note that $T_N$ is not simply related to $J$ due to the 2D character
of the magnetic interactions in cuprates\cite{rossat}.}
\label{table1}
\end{table}

\clearpage

\clearpage


\begin{thebibliography}{99}
\bibitem{formfactor} We used an anisotropic form factor which is 
more consistent with our data, in agreement with the copper form factor 
found in YBCO\cite{shamoto}.
\end{thebibliography}
\end{document}